\documentclass[twocolumn,aps,prc,showpacs]{revtex4}
\usepackage{amsmath,bm}
\usepackage{graphicx}
\begin{document}

\draft

\title{Elliptic flow of $\phi$ mesons and strange quark collectivity at RHIC}

\author{J. H. Chen$^{a,b}$}
\author{Y. G. Ma$^{a}$} \thanks{Corresponding author. Email: ygma@sinap.ac.cn}
\author{G. L. Ma$^{a,b}$}
\author{X. Z. Cai$^{a}$}
\author{Z. J. He$^{a}$}
\author{H. Z. Huang$^{c}$}
\author{J. L. Long$^{a}$}
\author{W. Q. Shen$^{a}$}
\author{C. Zhong$^{a}$}
\author{J. X. Zuo$^{a,b}$}

\affiliation{$^a$Shanghai Institute of Applied Physics, Chinese
Academy of Sciences, Shanghai 201800, China}
\affiliation{$^b$Graduate School of the Chinese Academy of
Sciences, Beijing 100080, China} \affiliation{$^c$University of
California, Los Angeles, CA90095, USA}
\date{\today}
\begin{abstract}

Based on A Multi-Phase Transport (AMPT) model, we have studied the
elliptic flow $v_{2}$ of $\phi$ mesons from reconstructed
$K^{+}K^{-}$ decay channel at the top Relativistic Heavy Ion
Collider energy at Brookhaven National Laboratory. The dependences
of $v_{2}$ on transverse momentum $p_T$ and collision centrality
are presented and the rescattering effect of $\phi$ mesons in the
hadronic phase is also investigated. The results show that
experimental measurement of $v_{2}$ for $\phi$ mesons can retain
the early collision information before $\phi$ decays and that the
$\phi$ $v_2$ value obeys the constituent quark number scaling
which has been observed for other mesons and baryons. Our study
indicates that the $\phi$ $v_2$ mostly reflects partonic level
collectivity developed during the early stage of the
nucleus-nucleus collision and the strange and light up/down quarks
have developed similar angular anistropy properties at the
hadronization.

\end{abstract}



\maketitle
\section{ Introduction}

Elliptic flow in heavy ion collisions is a measure of the
azimuthal angular anisotropy of particle distribution in momentum
space with respect to the reaction plane~\cite{Olli}. The
magnitude of the eeliptic flow depends on both initial spatial
asymmetry in non-central collisions and the subsequent collective
interactions. The elliptic flow is thus sensitive to the
properties of the dense matter formed during the initial stage of
heavy ion collision \cite{Sorge,Danielewice,Kolb,Ran} and parton
dynamics \cite{ZhangB} at Relativistic Heavy Ion Collider (RHIC)
energies. Experimentally, elliptic flow has been measured as
functions of collision centrality, transverse momentum,
(pesudo)rapidity and particle species
\cite{STAR1,PHENIX1,PHOBOS,STAR1b,STAR2} in $^{197}$Au +
$^{197}$Au collisions from RHIC at Brookhaven National Laboratory
(BNL). The experimental results of charged kaon, proton and pion
\cite{PHENIX1} show that the elliptic flow first increases with
particle transverse momentum following the hydrodynamic behavior
and then reaches a saturation in intermediate transverse momentum
region. More importantly, a Number-of-Constituent-Quark (NCQ)
scaling has been discovered for identified particle elliptic flow
in the intermediate $p_T$ region for baryons and mesons. The data
of pion elliptic flow $v_2$ are somewhat higher than the NCQ
scaling, which can be attributed to the large contribution to the
pion yield from secondary decays~\cite{Resonan-Pion,Dong}. The NCQ
scaling at RHIC is an important indication for the effective
constituent quark degree of freedom at hadronization and the
formation of hadrons through parton coalescence
mechanism~\cite{Lin2,Fries}. Similarly, a Number-of-Nucleon
scaling for anisotropic flow of light nuclear clusters in nuclear
collisions at Fermi energies was recently demonstrated by Ma et
al.~\cite{Ma06}, which is interpreted as a consequence of nucleon
coalescence mechanism.

Strange quark dynamics is a useful probe of the dense matter
created at RHIC. Enhanced strangeness production \cite{Rafelski}
has been proposed as an important signal for the formation of the
Quark-Gluon Plasma (QGP) in nuclear collisions. The dominant
production of $s\overline{s}$ pairs via gluon-gluon interaction
may lead to a strangeness (chemical and flavor) equilibration time
comparable to the lifetime of the QGP whereas the strangeness
equilibration time in a hadronic fireball is much longer than the
lifetime of the hadronic fireball. The subsequent hadronization of
the QGP is then expected to result in an enhanced production of
strange particles. In particular, it has been argued that with the
formation of QGP the production of $\phi$ mesons is enhanced.
Furthermore $\phi$ mesons could retain the information on the
condition of the hot plasma at hadronization because $\phi$ mesons
interact weakly in the hadronic matter~\cite{Shor}. The
measurement of $\phi$ mesons has been of great interest in the
study of collision dynamics and the properties of the dense matter
created at RHIC~\cite{STAR,PHENIX,DUKE-phi}.

We use A Multi-Phase Transport (AMPT) model to investigate effect
of parton dynamics related to $\phi$ mesons. The AMPT consists of
four main components~\cite{AMPT}: initial conditions, partonic
interactions, conversion from partonic matter into hadronic matter
and hadronic interactions in collision evolution. The initial
conditions, which include the spatial distribution of participant
matter, minijet partons production and soft string excitations,
are obtained from the HIJING model~\cite{HIJING}. Scattering among
partons are modelled by Zhang's parton cascade (ZPC) \cite{ZPC},
which calculates two-body parton scatterings using cross sections
from pQCD with screening masses. In the default AMPT model
\cite{DAMPT} partons are recombined with their parent strings when
they stop interacting, and the resulting strings fragment into
hadrons according to the Lund string fragmentation model
\cite{Lund}. In the AMPT model with string melting scenario
\cite{SAMPT}, a quark coalescence model is used to combine parton
into hadrons. The evolution dynamics of the hadronic matter is
described by A Relativistic Transport (ART) model \cite{ART}.
Details of the AMPT model can be found in \cite{AMPT}.

In this paper, we present a detailed study of the elliptic flow of
$\phi$ mesons at the top RHIC energy based on the AMPT model with
the string melting scenario \cite{AMPT}. The string melting
scenario is believed to be much more appropriate than the default
AMPT scenario since the energy density in these collisions is much
higher than the critical density for the QCD phase transition. The
AMPT model with string melting scenario has successfully described
the elliptic flow of stable baryons and mesons \cite{AMPT,SAMPT}.
It can also describe higher-order anisotropic flow parameters
$v_n$ including the odd-n ones \cite{Chen}. In this work, we focus
on the $\phi$ mesons. We follow the method used in experimental
analysis by reconstructing $\phi$ mesons in the final state from
fitting the invariance mass distribution of all $K^{+}$ and
$K^{-}$ pairs with a Breit-Wigner function including the intrinsic
decay width of $\phi$ mesons \cite{PYTHIA}. The parton scattering
cross section is chosen as 10 mb. The transverse momentum and
collision centrality dependences of $\phi$ meson $v_{2}$ have been
studied and the NCQ-scaling has also been observed in the AMPT
model. In addition, the rescattering effect on $\phi$ meson
$v_{2}$ has been investigated in the hadronic phase using the ART
model where rescattering processes of $\phi$ mesons and their
decay daughters are included \cite{Pal}. We note that our study
based on the AMPT model for $\phi$ meson elliptic flow is
different from that of Ref. \cite{Chen-phi}, where dynamical quark
coalescence model has been used for studies of $\phi$ meson
production and elliptic flow.

\section{ Analysis Method}

\begin{figure}[htbp]
\includegraphics[scale=0.52,bb=60 60 520 220]{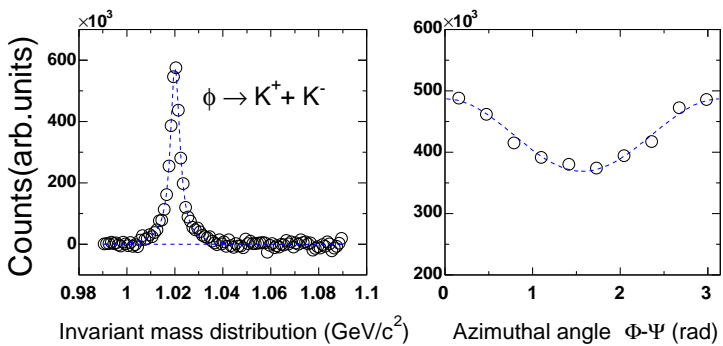}
\vspace{0.40truein} \caption{\footnotesize (Color online) Left
panel: $m_{inv}$ distribution from minimum bias (0-80\%) Au+Au
collisions at $\sqrt{s_{NN}}$ = 200 GeV. Dashed line is a fit of
Breit-Wigner function plus a linear background. Right panel:
azimuthal angular distributions of raw $\phi$ yields with respect
to the event plane. Dashed line represents the fit result. The
transverse momentum range of the figure is $0.4<p_{T}<3.0$
GeV/$c$.}\label{FIG.1}
\end{figure}

The reconstruction of $\phi$ mesons was accomplished by
calculating the invariant mass ($m_{inv}$), transverse momentum
($p_{T}$), azimuthal angle ($\Phi$) of pairs that formed from all
permutations of candidate $K^{+}$ with $K^{-}$ at a given $\phi$
rapidity range ($|y|<1.0$). The resulting $m_{inv}$ distributions
consist of the $\phi$ signal superimposed on a large background
that is predominantly combinatorial. The shape of the
combinatorial background was calculated using the mixed-event
method \cite{MIX}. A Breit-Wigner function plus a linear function
can describe the $m_{inv}$ distribution well. The left panel of
Fig.\ref{FIG.1} shows the $m_{inv}$ distribution for $\phi$ mesons
from minimum bias collisions using $\sim$ 1.8M Au+Au Monte Carlo
AMPT events.

\begin{figure}[htbp]
\includegraphics[scale=0.50,bb=60 60 500 380]{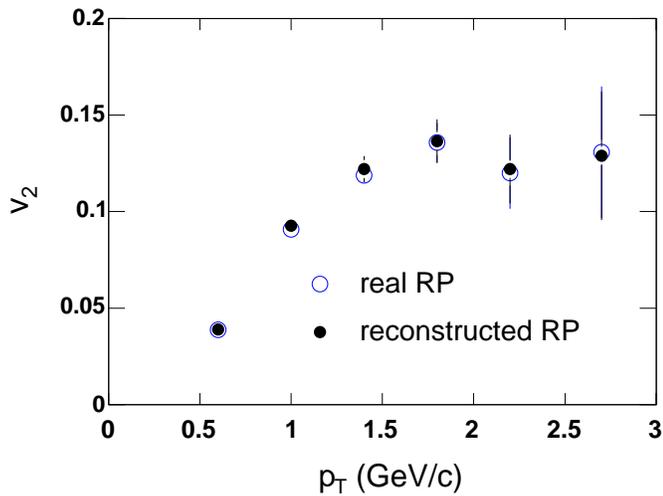}
\vspace{0.4truein} \caption{\footnotesize (Color online)
Comparison of elliptic flow values obtained from using real
reaction plane (RP) and event plane angle corrected for resolution
effect.} \label{FIG.2}
\end{figure}

The elliptic flow $v_{2}$ is calculated by the STAR standard
method \cite{STAR2} based on the distribution of particle raw
yields as a function of azimuthal angle $\Phi$ with respect to the
event plane angle $\Psi$. The raw yields of $\phi$ mesons are
extracted from fits to $m_{inv}$ distribution in each $p_T$ and
$\Phi-\Psi$ bin. The event plane angle $\Psi$ is determined from
the azimuthal distribution of charged tracks within a window of
$0.2<p_{T}<2.0$ GeV/$c$ and pseudo-rapidity $|\eta|<1.0$, which is
used as an estimate of the reaction plane angle
\cite{Danielewicz2,Poskanzer}. To avoid autocorrelations, tracks
associated with a $\phi$ candidate are explicitly excluded from
the event plane calculation. The right panel of Fig.\ref{FIG.1}
shows the azimuthal angular distribution of raw $\phi$ yields with
respect to the event plane from the minimum bias collisions in the
$0.4<p_{T}<3.0$ GeV/$c$ range. Dashed line is the fitting result
with a function $\frac{dN}{d(\Phi-\Psi)} = A
[1+2v_{2}cos2(\Phi-\Psi)]$, where $A$ is a normalization constant.
The finite resolution in the approximation of event plane as
reaction plane smears out the azimuthal angular distribution and
leads to a lower value in the apparent anisotropy parameters
\cite{Poskanzer}. This event plane resolution is determined by
dividing each event into random sub-events and calculating the
difference in event plane angles between sub-events. We obtained
an event plane resolution of 0.91, which is $20\%$ larger than the
experimental resolution reported by STAR \cite{STAR2}. This is due
to the fact that the number of tracks per event used in our
simulation is larger than that from the data. In order to verify
our resolution correction, we have also calculated the $v_{2}$
with respect to the real reaction plane ($\Psi=0$) which is known
a priori in our model calculation. From Fig.~\ref{FIG.2}, the
$v_2$ extracted from the true reaction plane is in good agreement
with the one extracted from the event plane corrected for
resolution effect. Our result also illustrates that the
experimental elliptic flow analysis method can faithfully describe
the magnitude of the elliptic flow for $\phi$ mesons.

\section{ Results}

\begin{figure}
\vspace{-0.10truein}
\includegraphics[scale=0.50]{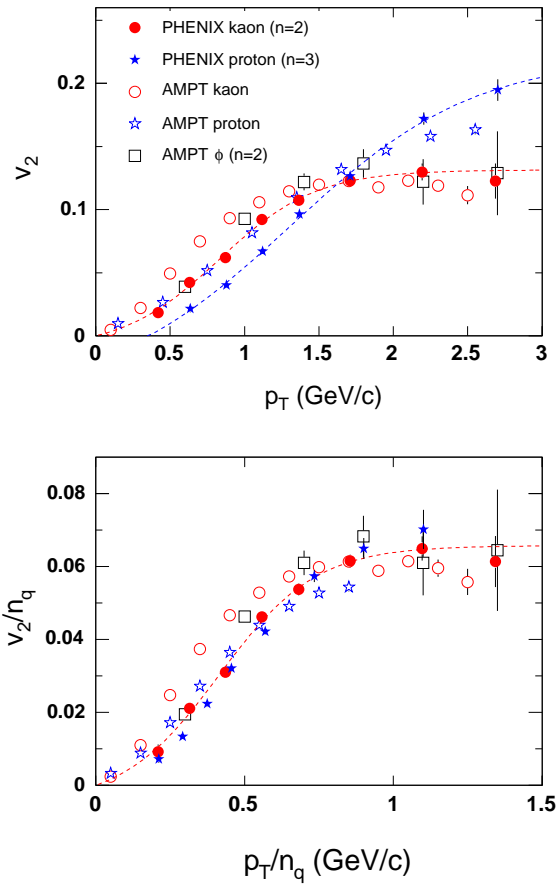}
\vspace{0.05truein} \caption{\footnotesize (Color online) Top
panel: $p_T$ dependence of the $v_2$ for $\phi$ mesons compared
with $K^{+}$+$K^{-}$ and $p+\overline{p}$. Data  are taken from
Ref.~\cite{PHENIX1}. Dot-dashed lines are the fitted results with
function ( $f_{v_{2}}$(n)
 = $\frac{a n}{1 + exp(-(p_T/n - b)/c)}$ - dn, where $a, b, c$ and $d$ are
the fit parameters, $n$ is the constituent-quark number ). Bottom
panel: NCQ scaled $v_{2}$ as a function of NCQ scaled transverse
momentum. The error bars represent statistical errors only.}
\label{FIG.3}
\end{figure}

The upper panel of Fig.\ref{FIG.3} shows the elliptic flow $v_{2}$
of $\phi$ mesons from minimum-bias AMPT $^{197}$Au + $^{197}$Au
collisions at $\sqrt{s_{NN}}$ = 200 GeV. Experimental data
\cite{PHENIX1} of $K^{+}$+$K^{-}$ and $p+\overline{p}$ are also
presented for comparison. We note that the $\phi$ meson $v_{2}$,
in compaison with $v_2$ of charged kaons and protons all from the
AMPT model, satisfies the mass ordering behavior of $v_2$
predicted by the hydrodynamic model calculation \cite{Huovinen} in
the low $p_T$ region. However, the AMPT calculation results of
$v_2$ for charged kaons and protons are about 25$\%$ larger than
the experimental data in the $p_{T}$ $<$ 1.5GeV/c region. This may
result from the large parton scattering cross section (10 mb) used
in this Monte Carlo AMPT calculation. In order to explore the
intermediate $p_T$ (1.5$<p_T< $4.0 GeV/$c$) phenomenon, where the
so-called NCQ-scaling in elliptic flow for identified particles
has been observed at RHIC \cite{STAR1} and the quark coalescence
or recombination mechanism has been used to explain the scaling
\cite{Lin2,Fries}, the large parton scattering cross section is
needed in order to produce the magnitude of the elliptic flow in
the intermediate $p_T$ region matching the experimental
measurement \cite{SAMPT}.

Our AMPT calculation indicates that the $v_{2}$ of $\phi$ mesons
at the intermediate $p_T$ seems to saturate and to follow the same
behavior as that of $K^{+}$+$K^{-}$. The lower panel of
Fig.~\ref{FIG.3} shows elliptic flow $v_{2}$ normalized by the
number of constituent quarks for charged kaons, protons and $\phi$
mesons.  In the intermediate $p_T$ region of $p_T$/$n_{q}$ $>$ 0.6
GeV/$c$ the elliptic flow of charged kaons, protons and $\phi$
meons from the AMPT calculation seems to satisfy the NCQ scaling.
This result implies that u, d and s quarks in the initial partonic
matter formed in relativistic heavy ion collisions develop
significant collectivity with strength characterized by
$v_{2}/n_{q}$.

\begin{figure}[htbp]
\includegraphics[scale=0.50,bb=60 60 500 400]{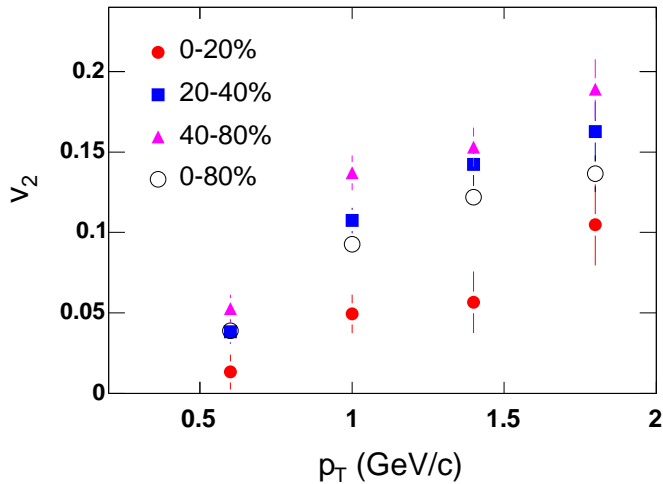}
\vspace{0.400truein} \caption{\footnotesize (Color online) $v_{2}$
of $\phi$ meson as a function of $p_T$ in the centralities of
0-20$\%$, 20-40$\%$, 40-80$\%$ and 0-80$\%$.  The error bars
represent statistical errors only.} \label{FIG.4}
\end{figure}

\begin{figure}
\includegraphics[scale=0.50]{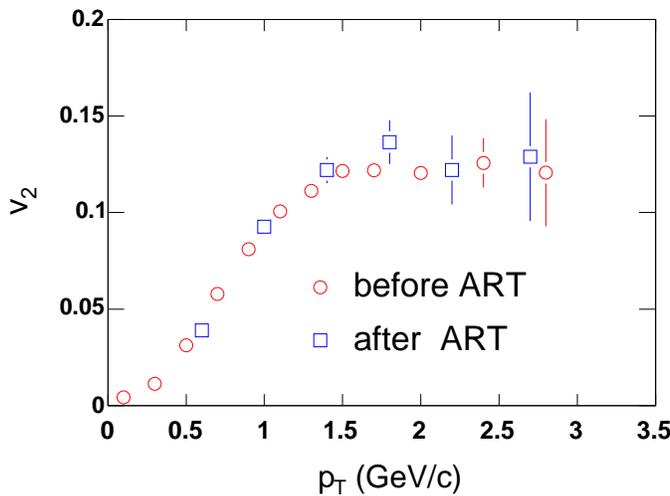}
\vspace{-0.1truein} \caption{\footnotesize (Color online) The
hadronic rescattering effect on elliptic flow $v_{2}$ of
$\phi$-meson from AMPT model with string melting scenario.}
\label{FIG.5}
\end{figure}

We have also studied the collision centrality dependence of $\phi$
meson $v_{2}$ at several centrality intervals: 0-20$\%$, 20-40$\%$
and 40-80$\%$ as well as $0-80\%$. Fig.~\ref{FIG.4} shows that for
each centrality bin, $v_{2}$($p_T$) of $\phi$ mesons increases
with their $p_T$. Among different centrality bins, the values of
$v_{2}$($p_T$) increase from central to semi-peripheral
collisions, which can be understood as a result of increasing
initial spatial eccentricity.

As the $K^{+}+K^{-}$ pair from the decay of a $\phi$ meson is
likely to undergo rescatterings in the medium during the hadronic
evolution, this might lead to a reconstructed $K^{+}K^{-}$
invariant mass situated outside the original $\phi$ meson mass
peak. It is thus of interest to study the in-medium rescattering
effect in details. This is carried out by turning on and off the
ART process during the hadronic evolution in the AMPT model. The
elliptic flow $v_2$ of $\phi$ mesons without ART indicates the
elliptic flow developed before the hardonic rescattering stage. In
contrast, the $v_2$ of $\phi$ mesons after ART includes all
contributions from both partonic and hadronic stages. In
Fig.~\ref{FIG.5}, the $v_{2}$ for $\phi$ mesons before ART are
directly extracted from the AMPT model without the ART processes.
In that case, $\phi$ mesons are explicitly present and do not need
to be reconstructed from $K^{+}+K^{-}$ decay channels. The $v_2$
after the ART processes is reconstructed from $K^{+}+K^{-}$ pairs
and the hadronic rescatterings are mostly due to kaon
rescatterings in the hadronic stage. Error bars are statistical
errors only. An interesting feature in Fig.~\ref{FIG.5} is that
the two scenarios are in good agreement with each other after
$p_T$ $>$ 0.4 GeV/c. In this case, the final hadronic rescattering
effect on the $\phi$ meson elliptic flow can be ignored within the
errors. Our study of $\phi$ $v_2$ confirms that $\phi$ mesons can
retain useful information from the early stage of the nuclear
collisions.

\section{ Summary}

In summary, we have presented a study of elliptic flow of
$\phi$ mesons using reconstucted $K^{+}K^{-}$ pairs
from minimum-bias $^{197}$Au + $^{197}$Au collisions
at $\sqrt{s_{NN}}$ = 200 GeV in a multi-phase transport model with
string melting scenario. The $v_{2}$ of $\phi$ mesons seems to
exhibit a similar behavior as other mesons. A NCQ-scaling phenomenon
of elliptic flow has been observed for $\phi$ mesons from the
reconstruction of $K^+ K^-$ pairs. The coefficient
$v_{2}$($p_T/n_{q}$) of $\phi$ mesons represents essentially the
momentum space anisotropy of constituent strange quarks that have
arisen from the partonic collectivity developed in the initial
stage of heavy ion collisions. The collision centrality dependence
of elliptic flow for $\phi$ mesons has also been studied. It is
found that the $\phi$ meson elliptic flow increases from central
to semi-peripheral collisions as a result of increasing initial
spatial eccentricity. We have also studied the in-medium hadronic
rescattering effect on elliptic flow of $\phi$ mesons. The results
confirm that within error bars, our reconstructed $\phi$ meson
$v_2$ can retain the early information before it decays. Comparing
our predictions with the RHIC data for the elliptic flow of $\phi$
mesons is expected to shed light on these issues.


We would like to acknowledge Dr. C. M. Ko, Z. W. Lin, B. Zhang and
B. A. Li for using their AMPT model. We also appreciate for
discussions and communications with Dr. C. M. Ko, N. Xu, and J. G.
Ma. This work was supported in part by the Shanghai Development
Foundation for Science and Technology under Grant Numbers
05XD14021 and 06JC14082, the  National Natural Science Foundation
of China under Grant No 10535010, 10328259  and 10135030.

{}
\end{document}